\documentclass[twocolumn,twoside,slac_two]{revtex4}
\usepackage{subfigure}
\usepackage{graphicx}
\usepackage{fancyhdr}
\usepackage{graphics}
\usepackage{epstopdf}
\usepackage{textpos}
\begin{document}

%%%%%%%%%%%%%%%%%%%%%% WRITE THE TITLE HERE %%%%%%%%%%%%%%%%%%%
\title{\centering Standard Model $H\rightarrow ZZ^{(*)}\rightarrow \ell\ell\ell\ell$, $\ell\ell\nu\nu$, $\ell\ell qq$ Searches with ATLAS at the LHC}
%%%%%%%%%%%%%%%%%%%%%% WRITE THE AUTHOR HERE %%%%%%%%%%%%%%%%%
\author{
\centering
\begin{center}
R. Rios, on behalf of the ATLAS Collaboration
\end{center}}
\affiliation{\centering Southern Methodist University, TX, 75205, USA}
%%%%%%%%%%%%%%%%%%%%%% WRITE THE ABSTRACT HERE %%%%%%%%%%%%%%%%
\begin{abstract}
The latest results on the direct experimental search for a Standard Model Higgs boson decaying to a pair of Z bosons are presented.  Three distinct final states are considered: $H\rightarrow ZZ^{(*)}\rightarrow \ell\ell\ell\ell$, $H\rightarrow ZZ\rightarrow \ell\ell\nu\nu$, and $H\rightarrow ZZ\rightarrow \ell\ell qq$, where $\ell = e, \mu$.  A dataset of more than 1fb$^{-1}$ of proton-proton collisions at $\sqrt{s} = 7$ TeV collected by the ATLAS detector at the CERN LHC during the on-going 2011 run is used.  In this model and for these final states, Higgs boson masses in the ranges 192 GeV $< m_H < $ 196 GeV, 214 $ < m_H < $ 222 GeV, and 340 $< m_H < $ 460 GeV have been excluded at the 95\% confidence level.
\end{abstract}

\maketitle
\thispagestyle{fancy}


\section{Introduction}
\begin{figure}
\includegraphics[scale=0.35]{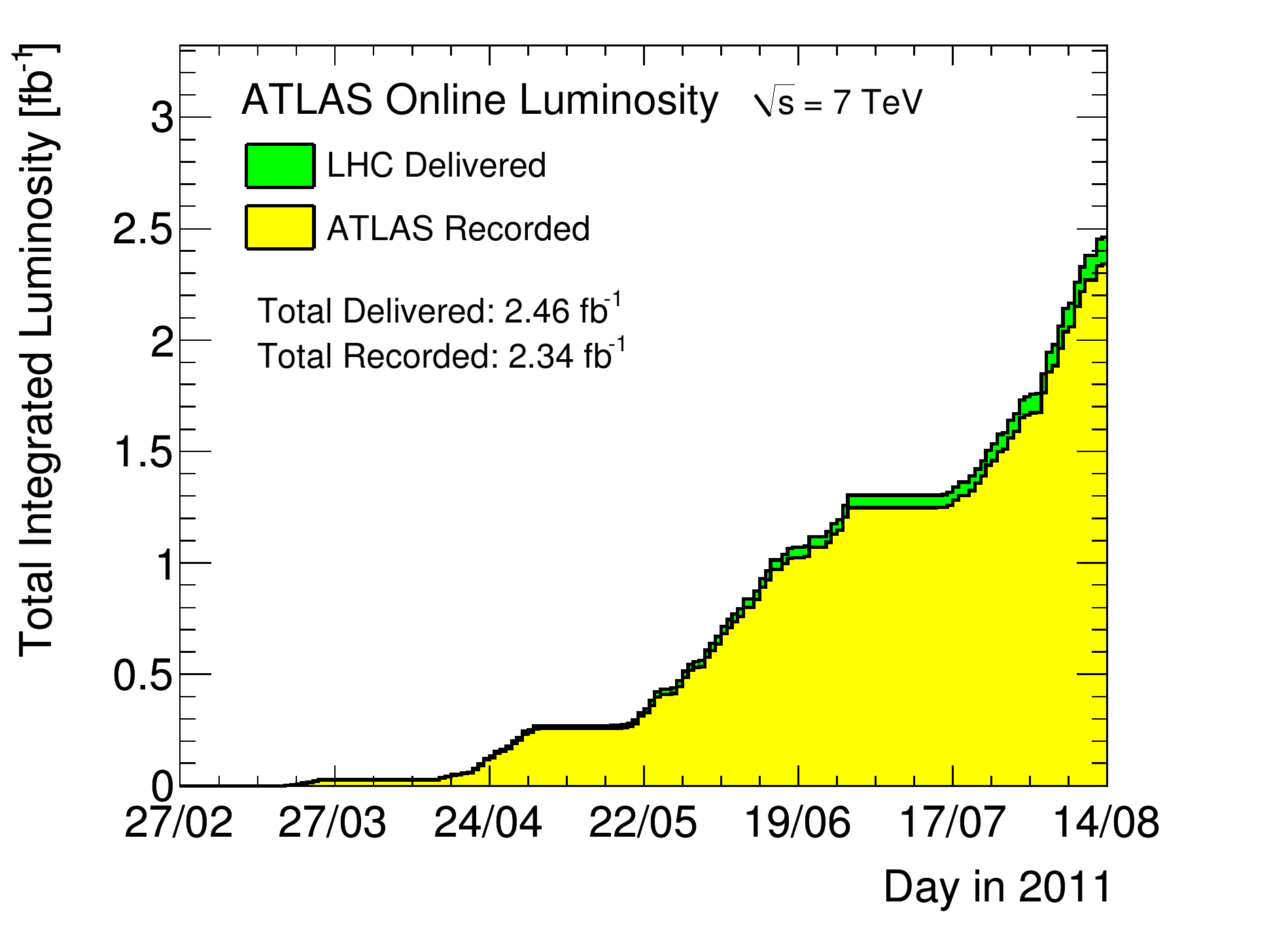}
\caption{Total integrated luminosity recorded by the ATLAS experiment by mid-August, 2011.}
\label{fig:lumi1011}
\end{figure}

One of the top quests of science today is to understand the mechanism responsible for electroweak symmetry breaking and the generation of mass in elementary particles.  The search for the Higgs boson, one possible solution to this quest, is one of the most important goals of the Large Hadron Collider (LHC) physics program and for the ATLAS (A Toroidal LHC ApparatuS) detector-experiment at CERN \cite{ATLASdetector}.
Direct searches by CERN's LEP collider experiments and the CDF and D0 experiments at the Fermilab Tevatron have resulted in Higgs boson exclusion limits for Higgs masses lower than 114.4 GeV and between 158 GeV and 175 GeV [\cite{LEP_results}, \cite{Tevatron_results}].
Since 2010, the LHC has been delivering proton-proton collisions at a center of mass energy of 7
TeV corresponding to an integrated luminosity of 48 pb$^{-1}$ (in 2010) and 2.55 fb$^{-1}$ (by mid-August 2011) collected by the ATLAS 
experiment (Figure~\ref{fig:lumi1011}). These data have been used to search for a Standard Model Higgs boson in a variety of final states.

A summary of the sensitivity for three distinct final state channels are considered in this paper, $H\rightarrow ZZ^{(*)}\rightarrow \ell\ell\ell\ell$, $H\rightarrow ZZ\rightarrow \ell\ell\nu\nu$, and $H\rightarrow ZZ\rightarrow \ell\ell qq$, where $\ell = e, \mu$.  As the direct search for Higgs bosons in data have not been observed, upper limits are set on the Higgs boson cross section relative to the Standard Model cross section at the 95\% CL using the CL$_s$ modified frequentist formalism \cite{freq_form} with the profile likelihood test statistic \cite{test_stat}.

\section{The ATLAS Experiment at the LHC}
The ATLAS detector is one of two multi-purpose detectors at the LHC; it is built with a forward-backward symmetric cylindrical geometry and designed to study high energy proton-proton collisions.  The Inner  Detector (ID) consists of a silicon pixel detector, a silicon microstrip detector, and a transition radiation tracker.  The ID is surrounded by a thin superconducting solenoid, which produces a 2 T magnetic field.  A high-granularity liquid argon (LAr) sampling calorimeter measures the energy and position of electromagnetic showers.  An iron-scintillator tile calorimeter provides hadronic coverage in the central rapidity range.  The forward rapidity regions are instrumented with LAr calorimetry for both electromagnetic and hadronic measurements.  The muon spectrometer surrounds the calorimeters and consists of three large superconducting toroids, each with eight coils, a system of precision tracking chambers, and detectors for triggering.  A three-level trigger system selects events to record for use with offline analyses.

\section{Results}
At the LHC, the Higgs boson is produced via several processes, among which, the most important are gluon-gluon fusion, vector-boson fusion, and associated production.  Gluon-gluon fusion is the dominant cross section for Higgs boson masses up to 1 TeV.  Higgs decay depeneds on its mass.  In the following sections, different channels are presented for differing masses.

\subsection{$H\rightarrow ZZ^{(*)}\rightarrow \ell\ell\ell\ell$}
\subsubsection{Event Selection}
Standard Model Higgs decay into four leptons is characterized by a final state, which consists of two, same flavour and oppositely charged, isolated lepton pairs.  The Higgs boson searches for this final state was performed with electrons and muons for Higgs masses between 110 GeV and 600 GeV.  Depending on the mass of the Higgs boson, one expects that there is at least one di-lepton pair with an invariant mass that is consistent with the mass of the on-shell $Z$ boson.  Kinematic cuts are applied to suppress contributions from reducible backgrounds: $Z$+jets and $t\bar{t}$ both of which can be suppressed with isolation/impact parameter cuts; and an irreducible background: $ZZ\rightarrow4\ell$.
\subsubsection{Background Estimation}
The shape and normalization of the expected backgrounds have been confirmed by a comparison of MC simulated event samples and the observed event yield in the data. $ZZ$ production was normalized to MC expectation, including both $qq/gq\rightarrow ZZ$ and $gg\rightarrow ZZ$.  An integrated luminosity uncertainty of 3.7\% and theory uncertainty of 15\% is associated with the normalization.  Top production is normalized to MC epectation and verified in the control region, with an associated theory normalization uncertainty of 10\%.  Normalization of $Z$+jets production is done using  control regions where $Z(\rightarrow\ell\ell) + \mu\mu/ee$ without applying isolation or impact parameter requirements.  Separate components, like heavy flavour, fakes, electroweak are taken into consideration and the normalization is extrapolated to the signal region with an uncertainty of 20\% - 40\%.
\subsubsection{Exclusion Limit}
\begin{figure}
\includegraphics[scale=0.25]{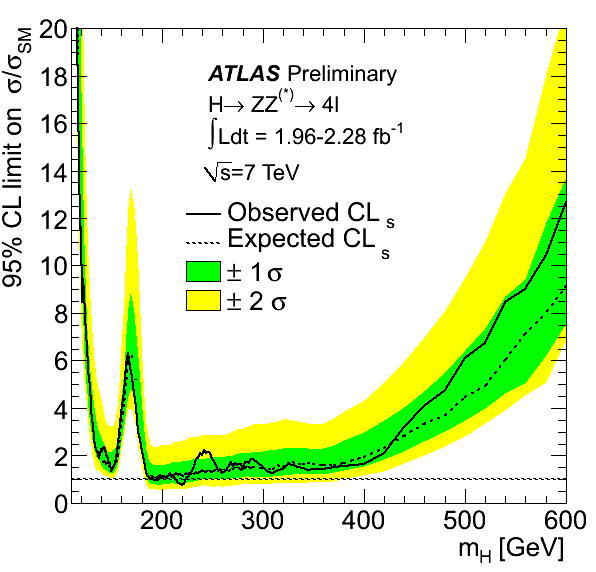}
\caption{Excluded signal cross section with respect to the SM rate at the 95\% CL for $H\rightarrow ZZ\rightarrow \ell\ell\ell\ell$.}
\label{fig:hzz4l}
\end{figure}

The search in this channel is sensitive to low and high Higgs masses with expected limit being at least 1.05 times the SM rate for a luminosity (dependent upon the final state) between 1.96 fb$^{-1}$ and 2.28 fb$^{-1}$.  Exclusion limits at the 95\% confidence level are set with respect to the SM cross section rate for Higgs masses in the regions from 192 GeV to 196 GeV and 214 GeV to 222 GeV (Figure~\ref{fig:hzz4l}). 

\subsection{$H\rightarrow ZZ\rightarrow \ell\ell\nu\nu$}
\subsubsection{Event Selection}
Higgs searches were also performed in the $\ell\ell\nu\nu$ channel for masses between 200 GeV and 600 GeV.  In this channel, the signal is characterized by one pair of oppositely charged, same flavour leptons with an invariant mass consistent with the on-shell $Z$ boson and a large missing transverse energy (MET) from the two neutrinos in the final state. Backgrounds in this final state include top/di-boson production and $H\rightarrow WW\rightarrow \ell\nu\ell\nu$, an orthogonal final-state Higgs search channel. Some of the cuts applied to reduce contributions from background include MET requirements, azimuthal angle separation between combinations of leptons, $Z$, MET, or jets. The final discriminant used is the transverse mass: $m^{2}_{T}\equiv\left[\sqrt{m^{2}_{Z}+|\vec{P}^{\ell\ell}_{T}|^{2}}+\sqrt{m^{2}_{Z}+|\vec{P}^{miss}_{T}|^{2}}\right]^{2} - \left[\vec{P}^{\ell\ell}_T+\vec{P}^{miss}_{T}\right]^{2}$.
\subsubsection{Background Estimation}
Comparison of MC simulated event samples with the observed event yield in the data show no deviations from the SM background-only expectations. Di-boson backgrounds were taken from MC and have an associated uncertainty of $\sim$10\%. The top background expectation was taken directly from data.  Data driven methods were used for $W$+jets background, taken from from same-sign lepton pairs, and QCD, which had generally small contributions.
\subsubsection{Exclusion Limit}
\begin{figure}
\includegraphics[width=7cm]{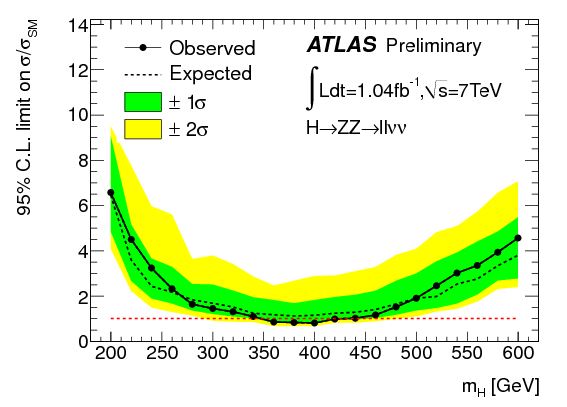}
\caption{Excluded signal cross section with respect to the SM rate at the 95\% CL for $H\rightarrow ZZ\rightarrow \ell\ell\nu\nu$.}
\label{fig:hzzllvv}
\end{figure}

This channel is sensitive in a wide mass range, with expected limits between 1.2 and 7 times the SM cross section rate for an accumulated luminosity of 1.04 fb$^{-1}$.  Exclusion limits with respect to the SM production rate are set at the 95$\%$ confidence level for Higgs masses from 340 GeV to 460 GeV. (Figure~\ref{fig:hzzllvv}).  

\subsection{$H\rightarrow ZZ\rightarrow \ell\ell qq$}
\subsubsection{Event Selection}
Higgs searches were performed with the $\ell\ell qq$ channel, where $\ell = e, \mu$ and were performed for Higgs masses between 200 GeV and 600 GeV.  Similar to the other channels, the signal in this channel is characterized by a same-flavour pair of isolated leptons with an invariant mass consistent with the on-shell $Z$ boson and a pair of jets whose invariant mass is also consistent with an on-shell $Z$ boson. Requirements on the number of b-tagged jets (b-tags), like azimuthal angle separation between two leptons or two jets, and high $p_{T}$ thresholds for leptons and jets are applied to reduce background contributions from $Z$+jets, $t\bar{t}$, and di-boson production. The final discriminant used in this channel is the $\ell\ell jj$ invariant mass where $m_{jj}$ is scaled to $m_Z$.
\subsubsection{Background Estimation}
Background estimation, like in other channels, is done with a mix of MC normalization and data-driven methods.  $Z$+jets is taken care of using the sidebands of $m_{\ell\ell}$ in bins of b-tags; an uncertainty of $<10\%$ is associated with un-tagged events (0, 1 b-jets) and $\sim20\%$ for tagged events (2 b-jets).  Top expectation is also taken from $m_{\ell\ell}$ sidebands and has an associated uncertainty of $\sim10\%$ from theory.  Di-boson backgrounds are taken from MC and have an associated uncertainty of $\sim10\%$.  The QCD background is estimated from data-driven methods - a background template is designed for $ee$ and a 2D sideband method using isolation and the invariant mass of the $\mu$ pair is used for $\mu\mu$, contributions are generally small.
\subsubsection{Exclusion Limit}
\begin{figure}[h]
\includegraphics[width=7cm]{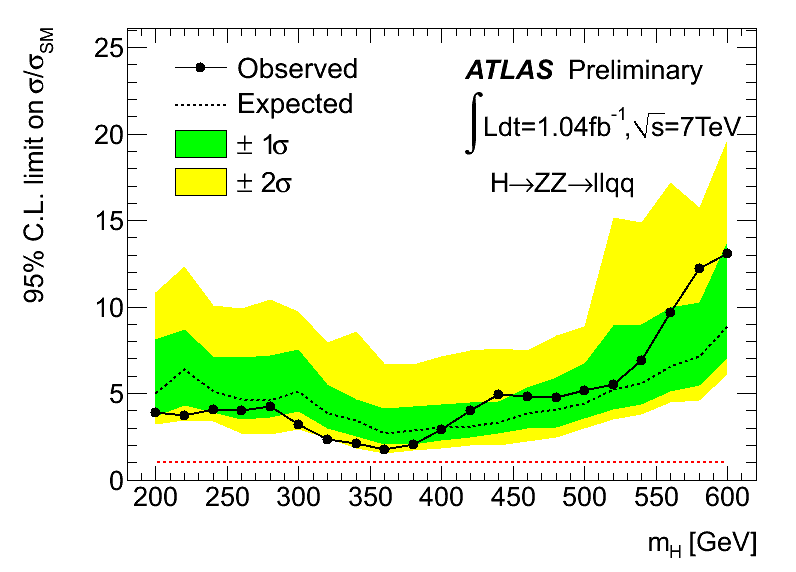}
\caption{Excluded signal cross section with respect to the SM rate at the 95\% CL for $H\rightarrow ZZ\rightarrow \ell\ell\nu\nu$.}
\label{fig:hzzllqq}
\end{figure}

This channel has good sensitivity in a wide mass range with expected limits between 2.7 and 9 times the SM cross section production rate, given a luminosity of 1.04 fb$^{-1}$.  An exclusion limit with respect to the SM production rate is set at the 95$\%$ confidence level for a Higgs mass of 360 GeV (Figure~\ref{fig:hzzllqq}).  

\vspace*{-.25cm}
\section{Summary}
The search for the Higgs decays: $H\rightarrow ZZ^{(*)}\rightarrow \ell\ell\ell\ell$, $H\rightarrow ZZ\rightarrow \ell\ell\nu\nu$, and $H\rightarrow ZZ\rightarrow \ell\ell qq$ has been presented using the data accumulated by the ATLAS detector during the 2011 run, which is up to 2.28 fb $^-1$, depending on the channel.  $H\rightarrow ZZ^{(*)}\rightarrow \ell\ell\ell\ell$ covers a large Higgs mass range, 100 Gev to 600 GeV and the $H\rightarrow ZZ\rightarrow \ell\ell\nu\nu$ and $\rightarrow \ell\ell qq$ channels complement $H\rightarrow ZZ^{(*)}\rightarrow \ell\ell\ell\ell$ by improving the sensitivity in the high mass region above 200 GeV.  The combined exclusion limit for Higgs masses in these three channels (for the given luminosities in each channel) is 192 GeV to 196 GeV, 214 GeV to 222 GeV, and 340 GeV to 460 GeV \cite{comb_note}.

%\begin{acknowledgments}
%The authors wish to thank JACoW for their guidance in preparing
%this template.
%
%Work supported by Department of Energy contract DE-AC03-76SF00515.
%\end{acknowledgments}

%% Create the reference section using BibTeX:

\end{document}